\documentclass{aa}
\usepackage{epsfig}

\newcommand{\teff}{T$_{\rm eff}$}
\newcommand{\nli}{$\log n$(Li)}
\newcommand{\nbe}{$\log n$(Be)}

\begin{document}

\title{UVES Be observations of early--G dwarfs in old clusters
\thanks{Based on observations collected at European Southern Observatory, Chile
(65.L-0427). Part of the data were obtained as part of an ESO Service Mode run.
}}

   \subtitle{ }

   \author{S. Randich\inst{1} \and F. Primas\inst{2} \and
         L. Pasquini\inst{2} \and R. Pallavicini\inst{3}}

   \offprints{S. Randich, email:randich@arcetri.astro.it}

\institute{INAF/Osservatorio Astrofisico di Arcetri, Largo E. Fermi 5,
             I-50125 Firenze, Italy
\and
European Southern Observatory
Karl Schwarzschild Strasse 2, D-85748 Garching bei M\"unchen, Germany
\and
INAF/Osservatorio Astronomico di Palermo, Piazza del
                  Parlamento 1, I-90134 Palermo, Italy}

\titlerunning{UVES Be observations of old clusters}
\date{Received Date: Accepted Date}

\abstract{We have obtained the first beryllium measurements of 
late F/early G--type
stars in the old open cluster \object{M67} (4.5~Gyr) and in the intermediate
age cluster \object{IC~4651} (1.7~Gyr). One member of the young 
cluster \object{IC~2391} ($\sim 50$~Myr) was also observed. Our sample stars
have effective temperatures within a range of $+30$ -- $+380$~K from the
solar temperature. All our 
sample stars, including the
Sun and the young cluster star have, within the errors, the same
Be abundance.
This result implies
that late  F/early G--type stars
undergo very little (if any) Be depletion during their main-sequence life-time.
Since these stars have undergone some Li depletion, our finding is indicative
of shallow mixing, i.e. of a mixing process that can transport surface
material deep enough for Li burning to occur, but not deep enough
for Be burning. As shown in previous studies, the Li vs. Be diagram 
is a powerful diagnostic of
stellar interiors. In this context,
we do not find any evidence of correlated
Li and Be depletion; furthermore, a comparison with various models
shows that the Be pattern of our sample stars 
is compatible only with models including gravity waves. This class of 
models, however, cannot reproduce the Li observations of M~67.
}
\maketitle
\section{Introduction}
The comparison between the chemical composition 
of  meteorites, assumed to be representative of the
primordial solar cloud, and the Sun
shows that lithium is 140 times underabundant
in the solar photosphere (Anders \& Grevesse \cite{grev}).
This result cannot be explained as due to the action of convection only
since, according to standard stellar models,
the convective zone of solar type stars on the main sequence (MS) 
does not reach deep enough layers to enable lithium
destruction (e.g., Pinsonneault \cite{pins97}).
Besides Li depletion in the Sun,
several pieces of observational evidence have been found indicating that,
in contrast to standard models, late F/early G--type
stars do deplete Li while on the MS (e.g., Jeffries \cite{jef00},
Pasquini \cite{pas00} and references therein), 
witnessing the action of an additional process that can transport material
from the surface to the stellar interior.
Whereas
several models have been developed, which
take into account a more complex physics, including mass loss, diffusion,
slow mixing driven by rotation or gravity waves
(Swenson \& Faulkner \cite{faulk}, Chaboyer et al. \cite{chab}, Deliyannis
\& Pinsonneault \cite{dp97}, Montalb\'an \& Schatzman \cite{ms00}),
these models are still very poorly constrained; in particular, the question
remains how deep in the stellar interior the mixing extends.

\begin{figure*}
\psfig{figure=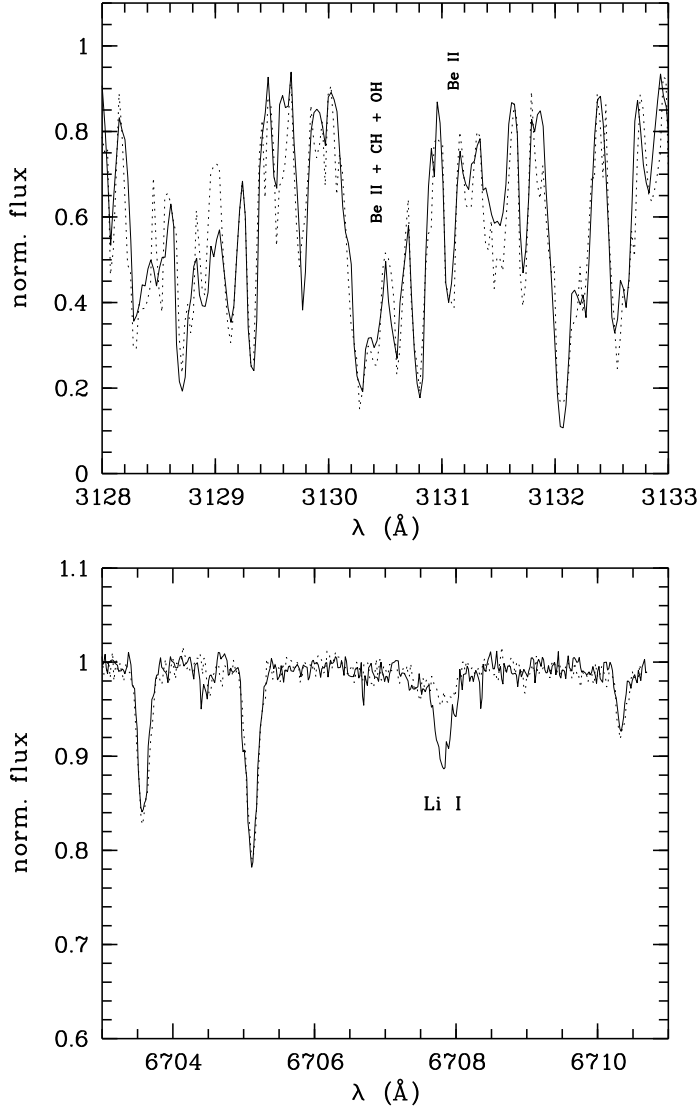, width=12cm, angle=0}
\caption{Comparison of the spectra in the Be (top panel)
and Li (bottom panel) spectral regions of the stars S1252 (solid line)
and S1256 (dashed line) in the old cluster M~67.
The figure clearly shows the lack of correlated Li and Be depletion;
whereas a large difference in Li strengths is evident, the spectra
in the Be region do not show a significant disagreement.
Li equivalent widths of 33 and 11~m\AA~were measured for S1252 and
S1256~respectively, resulting in a factor of $\simeq$ 5 difference
in Li abundance.}
\end{figure*}
\begin{table*}
\caption{Sample stars and Log of the observations}
\begin{tabular}{lllllcc}
 & & & & & & \\ \hline\hline
star & V & (B--V)$_0$ & Night & Exp. & S/N @ Be & S/N @ Li\\
 &  &  & (UT)& (sec) & &  \\
 & & & & &  &\\ 
$\alpha$ Cen A & $-0.1$  & 0.63  & 2000 Apr 11 & 4(b)/2(r) & 120 & 340 \\
IC~2391~SHJM2 & 10.3  & 0.57  & 2001 Feb 16 & 2400  & 30  & 210 \\
M~67~S988     & 13.18 & 0.534 & 2000 Apr 15 & 6300  & 40  & 120 \\
M~67~S994     & 13.18 & 0.535 & 2000 Apr 14 & 10800 & 50  & 160 \\
M~67~S1252    & 14.07 & 0.587 & 2000 Apr 16 & 5400  & 40  & 110 \\
M~67~S1256    & 13.67 & 0.595 & 2000 Apr 15 & 9100  & 45  & 120 \\
M~67~S969     & 14.18 & 0.622 & 2000 Apr 16 & 5400  & 40  & 110 \\
IC~4651 EG7   & 14.20 & 0.557  & 2000 Mar 26 & 14400 & 45  & 180 \\
IC~4651 EG45  & 14.20 & 0.568  & 2000 Mar 29 & 10800 & 40  & 140 \\
IC~4651 AT2105 & 14.02 & 0.545 & 2000 Mar 30 & 9540  & 40  & 140 \\
 & & & & &  &\\ 
 \multispan{6}{Numbering for M~67 from Sanders (\cite{sand}).\hfill}\\
 \multispan{6}{Numbering for IC~4651 from Eggen (\cite{eg}) and 
Anthony-Twarog et al. (\cite{at}).\hfill}\\
 & & & & & & \\ \hline\hline
\end{tabular}
\end{table*}
Since beryllium burns at temperatures $10^6$~K higher than Li
(Li and Be are destroyed by proton capture at the temperatures of
2.5 and 3.5~$\times 10^6$~K, respectively)
the simultaneous
determination of Li and Be abundances in the same star traces
stellar mixing to different depths, thus providing stringent
observational constraints on model predictions.
Whereas until recently it was commonly accepted that the Sun had
a Be abundance a factor of about two lower than meteoritic,
Balachandran \& Bell (\cite{bala}) carried out a revised analysis of
the solar spectrum in the near--UV and found a solar Be abundance 
\nbe$_{\odot}=1.40$, thus comparable 
with the meteoritic value \nbe$=1.42$
(Anders \& Grevesse \cite{grev}). They concluded
that Be is undepleted in the solar photosphere and suggested that
mixing in the solar interior is
shallower than previously thought. The result of Balachandran \& Bell
however is still controversial (e.g., King et al.
\cite{king}); in particular, the finding of little or no Be
depletion in the Sun cannot be regarded as a general conclusion,
since it relies on
absolute abundance measurements, which are hampered by large uncertainties.
Furthermore, we do not know whether and to what extent
our Sun is representative of the whole population of stars with similar
mass and evolutionary status.
Finally, according to most of the proposed models,
mixing depends on the whole evolutionary history of a star: hence,
knowledge of the Be abundance for only one star at the age of the Sun does
not allow discrimination between different models, since stars with
different parameter evolution (e.g. rotation) may undergo different
amounts of mixing.
Beryllium measurements in homogeneous samples of stars, such as stellar
clusters, with well
determined evolutionary status, chemical composition, and age,
are thus crucial. In addition, since most models
predict significant Be variations
on long timescales ($> 1$~Gyr), a meaningful comparison between model
predictions and observations should include old stars.

\begin{figure}
\psfig{figure=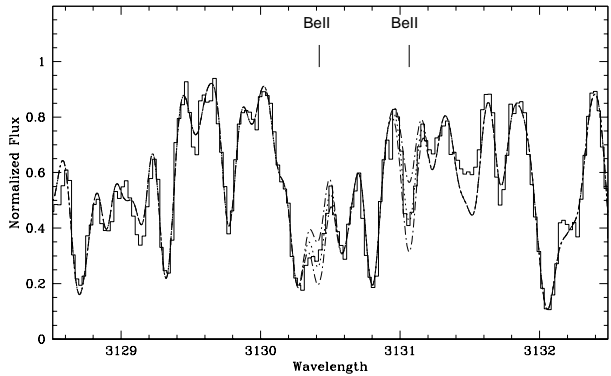, width=8.8cm, angle=0}
\caption{
The spectrum synthesis fit for the star S1252 in M~67.
The histogram represents the observed spectrum;
the dotted line denotes the best fit, while the two dot-dashed lines indicate
synthesis obtained with a factor of two lower/higher Be abundance.}
\end{figure}
\begin{figure}
\psfig{figure=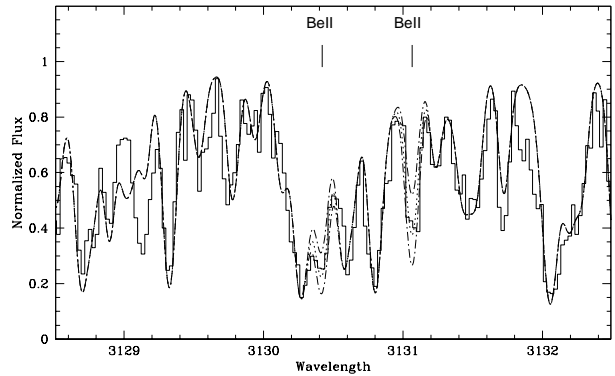, width=8.8cm, angle=0}
\caption{
Same as Fig.~2, but the spectrum synthesis for S1256 in M~67 is shown.}
\end{figure}
The Be feature that is best observed, the Be~{\sc ii} resonance doublet
($\lambda=3130.420$~\AA~and $\lambda=3131.064$~\AA), lies close
to the atmospheric UV cutoff; observations of Be are thus very
challenging and have so far been limited to the
brightest stars observable in the field  (e.g., Primas et al.
\cite{pri01}) and in the close-by Hyades
and Ursa Major clusters (Boesgaard \& Budge \cite{boe89}, Garc\'\i a
L\'opez et al. \cite{gar95}, Boesgaard \& King \cite{bk}). 
 
Thanks to the superior near-UV capabilities of UVES on VLT Kueyen (Dekker
et al. \cite{dek}),
we were able
to acquire for the first time high resolution spectra of the Be
region of 14$^{th}$ magnitude
G dwarfs in old clusters. 
In the present paper we report the results concerning Li and Be, while
an abundance analysis including heavy elements will be presented
in a forthcoming paper.
\begin{figure}
\psfig{figure=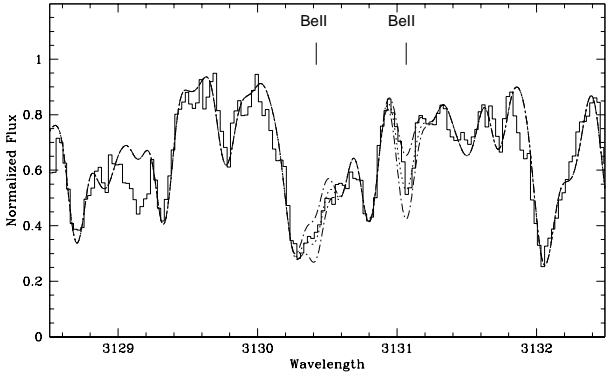, width=8.8cm, angle=0}
\caption{
Same as Fig.~2, but the spectrum synthesis for AT2105 in IC~4651 is shown.}
\end{figure}
\begin{figure}
\psfig{figure=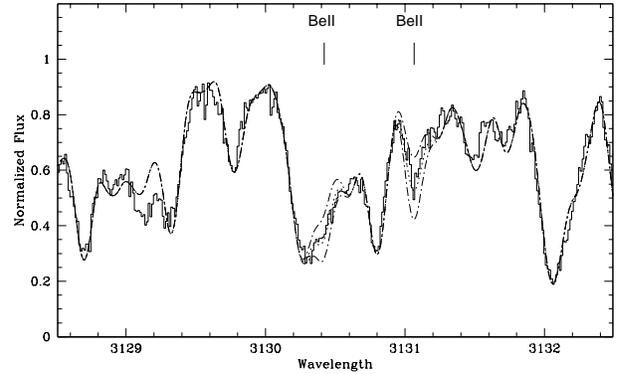, width=8.8cm, angle=0}
\caption{
Same as Fig.~2, but the spectrum synthesis for the star in the young clusters
IC~2391 is shown.}
\end{figure}
\begin{table*}
\caption{Effective temperatures, rotational velocities, and derived abundances}
\begin{tabular}{lccccc}
   & & & &  & \\ \hline\hline
star & \teff & v$\sin i$ & EW(Li) & \nli & \nbe \\
 &  (K)& (km/sec) & (m\AA) & \\
 & &   & &  &\\ 
$\alpha$ Cen A & 5800 & 2 & $9\pm 0.5 $ &  1.24 $\pm 0.11$ & $1.17\pm 0.11$\\
IC~2391~SHJM2 &  5970 & 9 & $150\pm 3$ & 3.45 $\pm 0.09$ & $1.11\pm 0.13$\\
M~67~S988     &  6153 & 6 & $12\pm 2 $ & 1.89 $\pm 0.13$ & $0.88\pm 0.13$\\
M~67~S994     &  6151 & 6 & $25\pm 2 $ & 2.27 $\pm 0.11$ & $1.16\pm 0.13$\\
M~67~S1252    &  5938 & 4 & $33\pm 3 $ & 2.35 $\pm 0.11$ & $1.11\pm 0.13$\\
M~67~S1256    &  5907 & 4 & $11\pm 1 $ & 1.60 $\pm 0.11$ & $1.16\pm 0.13$\\
M~67~S969     &  5800 & 4 & $28\pm 2 $ & 2.06 $\pm 0.10$ & $1.11\pm 0.13$ \\
IC~4651 EG7   &  6061 & 4 & $52\pm 3 $ & 2.61 $\pm 0.11$ & $1.11\pm 0.13$\\
IC~4651 EG45  &  6016 & 4 & $48\pm 3 $ & 2.52 $\pm 0.11$ & $1.16\pm 0.13$\\
IC~4651 AT2105 &  6110 & 8 & $67\pm 6$ &2.82 $\pm 0.12$ & $1.11\pm 0.13$\\ \hline\hline
 & &   & &  & \\ 
\end{tabular}
\end{table*}
\section{Observations}
Our sample includes five members of the solar age M~67 cluster (4.5~Gyr),
three members of the 1.7~Gyr old IC~4651,
the solar--analog $\alpha$
Cen A, and one member of the young (50~Myr)
cluster IC~2391. This star has just arrived on the zero age main
sequence and it is undepleted in Li (Randich et al. \cite{R01});  
under the fully
reasonable assumption that it is also undepleted in Be,
it can be used to set a zero point
to our abundance scale and it may provide an indication of the initial 
Be abundance. No known binaries are included in our sample.
The observations were carried out using UVES on VLT UT2 (Kueyen).
M~67 members and $\alpha$ Cen A were observed in Visitor Mode in April
2000; the young cluster member was also observed in Visitor Mode in
Feb. 2001, as part of a different program; finally, the three targets in
IC~4651 were instead observed in Service Mode during March 2000.
All the stars were observed using the same set-up of UVES; namely,
UVES was operated in Dichroic Mode using Cross Dispersers \#1 and \#3 in the 
Blue and Red arms, respectively. The Blue arm is equipped with a
EEV 2048$\times$4102 CCD, while the Red arm is equipped with a mosaic
of two CCDs made by a EEV 2048$\times$4102 CCD and a MIT-LL $2048\times 4102$
CCD.
Such a combination allowed us to
cover the spectral ranges from $\sim 3115 $ to 3940~\AA~in the blue and
from $\sim$ 4780 to 6810~\AA~in the red. The 15$\mu m$ pixels
and the use of a 1~arcsec wide
slit (projecting into 4 pixels)
and CCD binning $2\times 2$ and $1\times 1$ in the blue and
red, respectively, yielded
resolving powers R$\sim 40,000$ and R$\sim 45,000$.
The spectra were reduced using MIDAS and following
the usual steps: bias and inter-order background were subtracted from
both science and flat-field frames; the science frames were then optimally
extracted, flat-fielded, and wavelength calibrated. Finally, the extracted
orders were merged. The exposure times of cluster stars ranged between
40~min and 3~hrs; the sample stars were observed under different seeing 
conditions and variable
airmasses, due to the long exposure times. The final S/N ratios per 
spectral bin
are comprised
between $\sim 30$ and 50 in the Be region and between $\sim 100$ and 200 in the
Li region. A much higher S/N was achieved for $\alpha$ Cen A.
The sample stars together with
details on the observations are listed in Table~1.
In Fig.~1 we show the comparison
of the spectra of two M~67 members in the Be and Li spectral regions.
\section{Abundance and Error Analysis}
Effective temperatures were determined based on B--V colors using 
the same calibration employed in similar studies (Soderblom et al. 
\cite{sod93}). A surface gravity $\log g=4.44$ and microturbulence 
$\xi=1.1$~km/sec were assumed for all sample stars. 
Rotational velocities were retrieved from Pasquini et al. (in preparation)
for M~67 and IC~4651 and from Randich et al. (\cite{R01}) for the young
cluster member. \teff~and v$\sin i$ values are listed in Cols. 2 and 3
of Table~2.

Li and Be 
abundances were derived in a consistent way using Kurucz model 
atmospheres and codes (ATLAS9, SYNTHE, and WIDTH9 -- Kurucz \cite{kur}). 
Model atmospheres were computed for the exact stellar parameters of 
each star, by interpolating in the ATLAS9 grid of models officially 
released by Kurucz on CD-ROMs (Kurucz \cite{kur}). Solar abundances 
were taken from Anders \& Grevesse (\cite{grev}). 

Be abundances were determined by spectrum synthesis using the line 
list extensively tested by Primas et al. (\cite{pri97}). This compilation 
has been constrained by using a large sample of stars, and after having 
run several tests on the possible presence of unidentified lines in the Be~II 
doublet region. The main reason behind these tests (also carried out by 
similar studies) is the following: using the list of atomic and molecular lines 
derived from laboratory measurements only and model atmospheres currently 
available, it is very difficult to obtain a satisfactory fit of the 
$\lambda$3131.064~\AA~line with the solar photospheric abundance (\nbe$_{\odot}=
1.15$, as determined by Chmielewski et al. \cite{chm75}). Possible ways out 
include the introduction of one or more (still unidentified, hence 
``predicted'') lines or adjusting the oscillator strengths of neighbouring 
lines (cf. Garc\'\i a L\'opez et al. \cite{gar95}). Though different, these 
two approaches give similar results, a satisfactory fit of the solar 
spectrum with the abundance \nbe$_{\odot}=1.15$. For our analysis, we adopted 
the line list from Primas et al. (\cite{pri97}), including a predicted 
Fe~I line at 3131.043~\AA~, the strength of which had been constrained by 
using a large sample of stars of different \teff, $\log g$, and spanning 
three orders of magnitude in metallicity (cf. the original work for more 
details).   

We started our analysis from fitting the Kurucz Solar Flux Atlas (Kurucz et al. 
\cite{kur84}): using a solar model with \teff$_{\odot}=5777$~K, 
$\log g_{\odot}=4.437$, $\xi_{\odot}= 1.1$~km/sec, we obtained a solar Be 
abundance \nbe$_{\odot}=1.11$, in very good agreement with the value found 
by Chmielewski et al. (\cite{chm75}). To gain confidence on the 
robustness of our analytical method, we then fitted our own spectrum of the 
solar-analog $\alpha$ Cen A (\teff$=5800$~K, $\log g=4.44$, [M/H]=+0.1, cf. 
Primas et al. \cite{pri97}). We found \nbe$_{\odot}= 1.17$, in good agreement 
with the Sun and with values derived from previous analyses (within the 
quoted uncertainties - cf. Primas et al. \cite{pri97}, King et al. 
\cite{king}). Then, we applied the same technique to all our sample 
stars, including also the member of the young cluster IC~2391. 
Examples of some spectrum synthesis fits are shown in Figs.~2--5, while 
all the Be (derived by demanding a satisfactory fit of {\it both} Be lines) 
and Li abundances are reported in Table~2, together with their respective 
uncertainties. For the records, using model atmospheres computed with the 
overshooting option switched off (NOVER) translates into a systematic 
difference $\Delta$\nbe~$=-0.05$~dex for all stars. 

Li abundances were derived from the measured equivalent widths of the 
Li~{\sc i}~6707.8~\AA~feature (see Table 2). Given the resolution of
our spectra and the rotational velocity of the sample stars, the
Li feature was not blended with the nearby Fe~{\sc i}~6707.44~\AA~
line and, thus, we did not need to correct the measured Li EWs for the
contribution of the latter. For completeness and 
consistency, Li abundances for stars from previous studies (namely, Jones 
et al. \cite{jon99} for M~67 and Randich et al. \cite{r00} for IC~4651) 
have also been re-computed from the EWs published in the original works. 
We note that new abundances are in very good agreement with old ones. 

Errors in the derived abundances were computed from analyzing the dependence 
of Li and Be abundances to changes in the adopted stellar parameters. 
Uncertainties of $\pm 100$~K in \teff, 0.2~dex in $\log$~g, and 0.2~km/sec
in $\xi$ were assumed. The final uncertainty listed in Table~2 include also 
the sensitivity of Be to the placement of the continuum (estimated 
to be on the order of $\pm 3\%$) and of Li to errors in the measured EWs. 
There may be extra sources of systematic and random errors. 
In the case of Be, one may want to take into account extra uncertainties 
encountered during spectrum synthesis. During our analysis we have 
established that including one ``predicted'' line (or adjusting the 
oscillator strengths of one or more lines) implies a systematic difference 
of $\sim -0.05$~dex for our sample stars here (as to say, without the 
predicted line, one has to increase the Be abundance by +0.05~dex in order to 
get a similar satisfactory fit). Another way is to run a synthesis including 
the entire Kurucz list of ``predicted'' (as opposed to ``laboratory'') 
lines, in order to quantify the global effect on the continuum. Although 
related to the same problem, this may be considered a random source of error. 
We found it affects the continuum level by $\sim 3\%$, which corresponds to 
$\pm 0.05$~dex in beryllium. Another weak point to be aware of is the well 
known discrepancy between the solar photospheric Be (\nbe$_{\odot}=1.15$) and 
the abundance measured in meteorites (\nbe$_{\odot}=1.42$ determined by 
Anders \& Grevesse \cite{grev}). Though this may turn out to be only an 
apparent discrepancy (cf. Balachandran \& Bell \cite{bala}, who determined a 
solar photospheric Be in very good agreement with the meteoritic value by 
including new continuum opacity sources), we decided not to enter the 
discussion which, as mentioned above,
is still very much open, as also shown by Allende Prieto \& 
Lambert (\cite{all}) who find that there is no missing UV opacity. 

In conclusion, based on the quality of our observations and on the 
fact that our main 
conclusions will be based on differential comparison between stars 
characterized by similar parameters,
we believe that the {\it relative} error bars quoted in 
Table~2 are appropriate. In other words,
thanks to our choice of targets
of very similar nature and macroscopic parameters, we can safely compare
the relative light element abundances of our sample stars
without being affected by systematic effects.
This implies that differences in Be abundances are better constrained than
absolute abundances.
\section{Results and discussion}
\subsection{Be and Li abundances}
In Fig.~6 we show \nbe~vs. \teff~for our sample stars. The Sun
is also included in the figure.
Figure~6 and Table~2 clearly show that all our sample stars have,
within the errors, the same Be abundances: in particular, we find
for the Sun exactly the same abundance as for the 50~Myr
old IC~2391 member;
stars in M~67 (with the possible exception of S988 that shows a somewhat
lower abundance), IC~4651, and $\alpha$ Cen A also have
a similar abundance as the Sun. We obtain an average abundance \nbe
$=1.08 \pm 0.12$ for M~67 and \nbe~$=1.13 \pm 0.03$ for IC~4651.
We stress that the finding of similar abundances is extremely 
robust since it is based both on direct comparison
of the spectra and on a consistent
analysis of stars with very similar macroscopic parameters.
\begin{figure*}
\psfig{figure=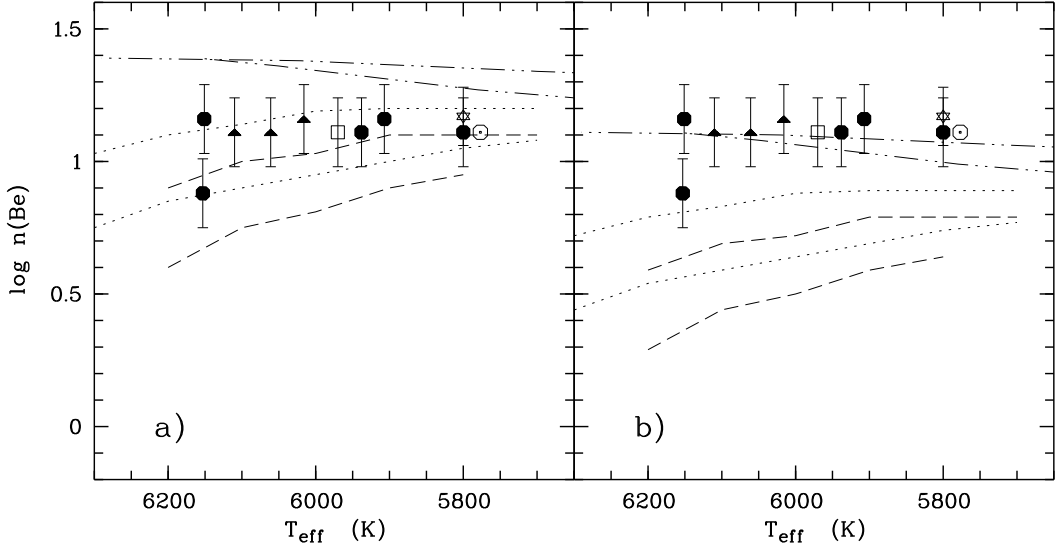, width=16.0cm, angle=-90}
\caption{
\nbe~vs. effective temperature for our sample stars.
Filled circles and triangles denote M~67 and
IC~4651 members, respectively, while the open square indicates
the young cluster member.
The Sun (Sun symbol) and $\alpha$ Cen A (star symbol)
are included in the figure. 
Curves denoting model predictions are shown in the figure. Dot-dashed
curves indicate the predictions of models including
gravity waves, 
with upper and lower curves denoting predictions at an age
of 1.7 and 4.0~Gyr, respectively.  
Dotted and dashed
curves indicate the predictions of rotationally induced mixing
at 1.7 and 4 Gyr; for each curve style, upper curves refer to models
with an initial rotational velocity of 10 km/sec, while lower 
curves denote the predictions of models
with an initial rotational velocity of 30 km/sec.
In panel a) theoretical curves
are normalized to an initial abundance equal to the meteoritic value 
\nbe$=1.42$, while in panel b) the curves are normalized to the
abundance \nbe$=1.11$ that we derived for the young cluster member.} 
\end{figure*}
\begin{figure*}
\psfig{figure=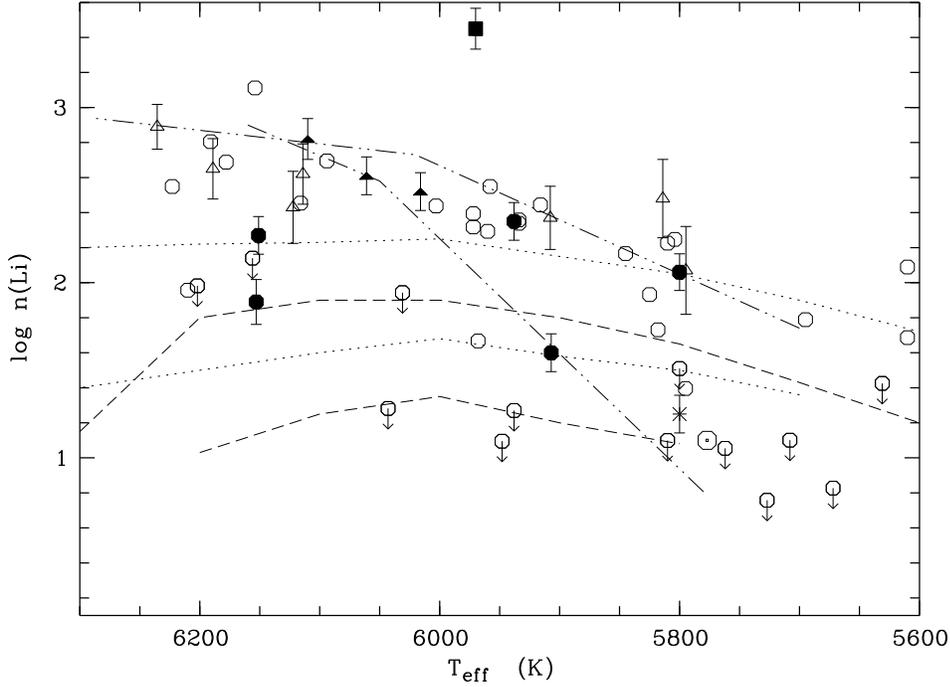, width=14.0cm, angle=-90}
\caption{
\nli~vs. effective temperature for our sample stars.
Symbols and curves are the same as in Fig.~6, but filled
symbols refer to present data (namely, filled circles: M~67; filled
triangles: IC~4651; filled square: young star; asterisk: $\alpha$ Cen A),
while open circles and triangles denote
Li abundances for M~67 and IC~4651 from previous studies
(Jones et al. \cite{jon99} and Randich et al. \cite{r00}).}
\end{figure*}
\begin{figure*}
\psfig{figure=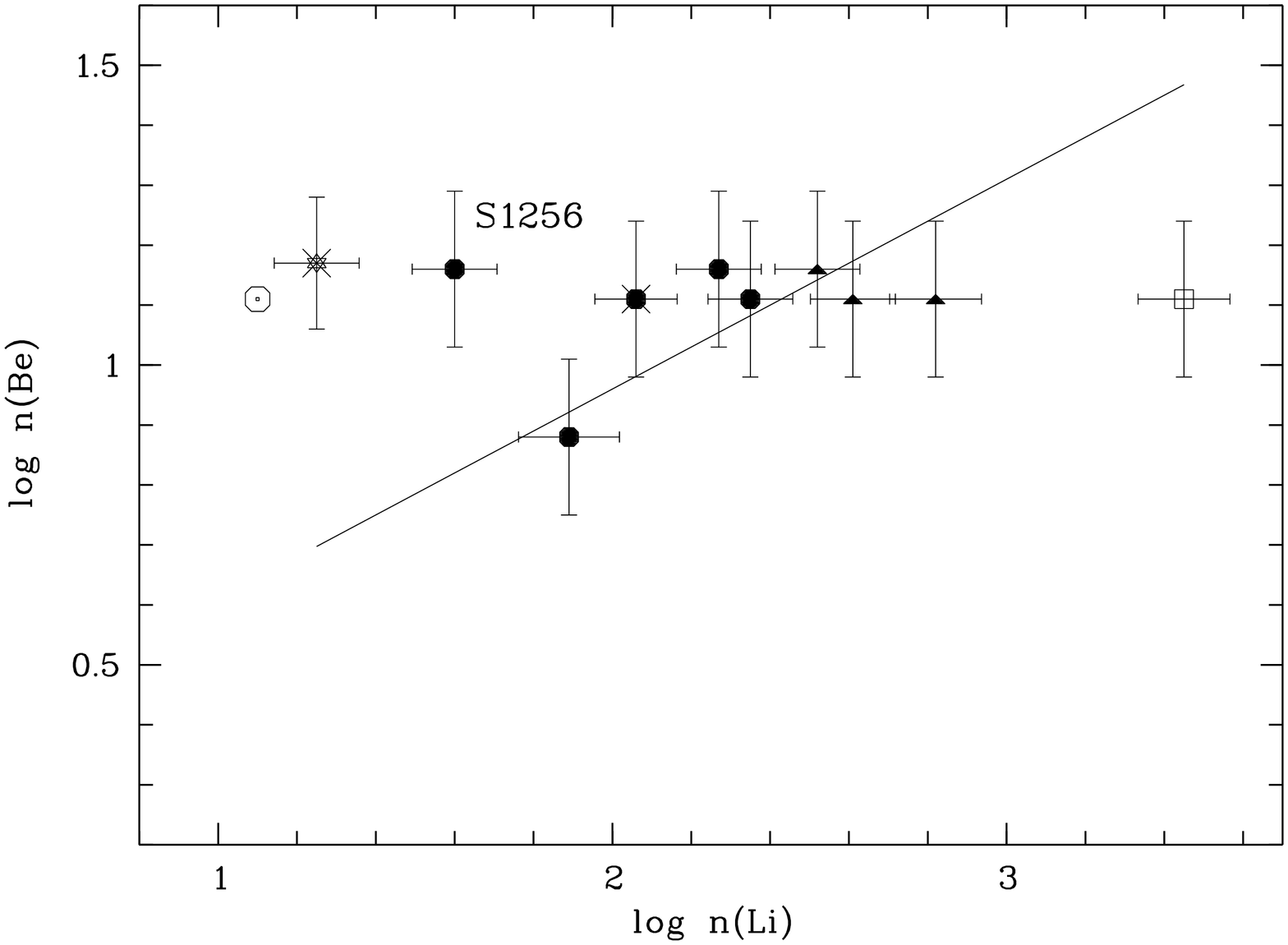, width=14.0cm, angle=-90}
\caption{\nbe~vs. \nli.
Circles and triangles
indicate M~67 and IC~4651 members, respectively;
the Sun is denoted by the Sun symbol,
$\alpha$ Cen A by a star symbol, and the young star in IC~2391 by
an open square. Crossed symbols denote stars with \teff~below
5850~K. 
The solid line indicate the best fit relationship
found by Boesgaard et al. (\cite{boe01}) for field stars with \teff~
in the interval 5850--6300~K. 
}
\end{figure*}
As we have mentioned, 
the Be abundance that we derive for the Sun is in agreement with
the value \nbe~$=1.15$ derived by Chmielewski et al. (\cite{chm75});
this value, as well the abundances of our sample stars,
are about a factor of two
lower than the meteoritic abundance \nbe~$=1.42$.
For the three clusters we have derived  similar, about solar, metallicity
(see also Randich et al. \cite{R01}): given the \nbe~vs. [Fe/H] relationship
(e.g., Boesgaard et al \cite{boe99}), this suggests
a similar initial Be abundance. 
Two hypotheses can be proposed to explain the difference between the 
abundances that we have derived for our sample stars
and the meteoritic Be abundance:
{\bf a)} All our sample stars, including the young cluster star,
have undergone a very similar amount of Be depletion; {\bf b)}
The difference between our best fit Be abundances and the meteoritic value
is simply due to 
different analytical methods adopted during the analyses. Under this hypothesis,
none of our sample stars has undergone any Be depletion, implying
that {\bf stars in this temperature range}
including the Sun, do not burn Be while on the MS. 
Hypothesis {\bf a)} would mean that some Be burning occurs
during the pre-main sequence (PMS) phases,
in contradiction with all
model predictions (e.g., Piau \& Turck-Chieze \cite{pt01}) and
inconsistently with the
observational evidence of no PMS Li depletion for stars in
this \teff~range (e.g., Randich et al. \cite{R01} and references therein);
in other words, the
simultaneous Be depletion and Li non--depletion in the young star would
be hardly explainable. For these reasons, we regard hypothesis {\bf a)}
as very unlikely.
Hypothesis {\bf b)} would instead be the proof of a shallow mixing process,
i.e. a mechanism that is
able to transport surface material down to the Li burning
layer, but not deep enough to destroy Be.

In Fig.~7 we plot \nli~vs. \teff~for our sample stars and
the stars studied by Jones et al. (\cite{jon99}) and
Randich et al. (\cite{r00}). 
The present study confirms the star-to-star scatter in Li
among M~67 members (Spite et al. \cite{spi87}, Garc\'\i a L\'opez
et al. \cite{gar88}, Pasquini et al. \cite{pas97}, Jones et al.
\cite{jon99}), indicating different amounts of Li depletion
for stars with the same mass and chemical composition. In agreement with
previous studies (Randich et al. \cite{r00}),
stars in the younger IC~4651 cluster 
do not show any major dispersion, suggesting that 
the mechanism responsible for the dispersion observed in M~67 acts at
relatively old ages.

\subsection{Be vs. Li depletion and comparison with the models}
As clearly pointed out by Deliyannis (\cite{deli00}), the \nbe~vs. \nli~diagram
is a very powerful tool to investigate the nature of stellar mixing and
to discriminate between different models. It is worth emphasizing
that in this diagram the shape of the relationship, i.e., the relative
Li and Be abundances, rather than absolute abundances, is providing most
of the information.
In Fig.~8 we show \nbe~vs. \nli~for our sample stars and the Sun;
the best fit relationship between \nbe~and \nli~ proposed
by Boesgaard et al. (\cite{boe01}) 
for field stars with $6300 \geq$ \teff~$\geq 5850$~K
is also shown in the figure.

Be surveys among field stars (e.g., Stephens et al. \cite{ste97}, Boesgaard
et al. \cite{boe01})
have reported the existence of a correlation between Li and Be depletion;
this correlation was interpreted as the proof of the action of slow mixing
driven by rotation; a similar result was found by Boesgaard \& King
(\cite{bk}) for Hyades dwarfs warmer than 5850~K.
At odds with the correlation found for field stars, and in contrast with the 
prediction of slow mixing induced by rotation (see e.g., Deliyannis \cite
{deli00}),
Fig.~8 does not show any clear
evidence of correlated Li and Be depletion; stars that have suffered
different amounts of Li destruction both in the same cluster and
in different clusters, have, within the errors,
the same Be content (see also Fig.~6). Our datapoints plotted in
Fig.~8 do not follow
the mean trend found for field stars with $6300 \geq$~\teff~$\geq 5850$~K.
This is not unexpected
for the stars in our sample with \teff~below 5850~K (namely,
the Sun, alpha Cen A, and S969 in M~67), for which 
no correlation between Li and Be is indeed found for the Hyades.
It is instead surprising that the warmest
stars in our sample do not follow the same relationship as field stars. 
Most of the stars in M~67 and IC~4651
are only marginally in agreement with the
relationship for field stars, with star S1256 in M~67 clearly
showing much less Be depletion than expected; viceversa, 
the young cluster star has a Be abundance significantly below the expected
value. If one changed the intercept of the curve to fit this star,
all the other sample stars would then lie much above the relationship.
We mention that different
temperatures were reported in the literature for S1256, a couple of which are
cooler than 5850~K. More in general,
our temperature scale is different from that of
Boesgaard et al. (\cite{boe01}) and
thus other stars in our sample, besides $\alpha$ Cen A, the Sun, and S969,
might fall outside the interval $5850 \leq$ \teff $\leq 6300$~K based
on the calibration of Boesgaard et al. 
In order to further investigate this point, we
re-determined the effective temperatures of field stars in the sample
of Boesgaard et al. using our calibration.
Whereas we do find differences as large as $\sim 250$~K,
we do not find a systematic offset and the range 5850--6300~K of
Boesgaard et al. reflects into an interval of 5810--6480~K in our scale.
In other words, given the uncertainties in the temperature scales,
the lower limit for the Li-Be correlation found for field stars cannot
be defined in an absolute way. On the other hand, the discrepancy between
the correlation found for field stars and the lack of correlation found
for our sample stars cannot be explained as due to different temperature
calibrations.

In the following, our results are compared with the predictions of various
models.

Models including diffusion actually do not warrant a detailed discussion: 
as discussed by Deliyannis (\cite{deli00}) or by Stephens et al.
(\cite{ste97}),
diffusion implies simultaneous Li and Be depletion and thus does not
fit the observed Be vs. Li diagram shown in Fig.~8.
In Figs.~6 and 7, instead, we show a quantitative comparison
with the predictions of models including slow mixing processes;
more specifically, we plot different curves denoting
theoretical predictions by models including slow rotationally
induced mixing (retrieved from Deliyannis \& Pinsonneault \cite{dp97}) 
and by
models including gravity waves (kindly provided by J. Montalb\'an --see
also Montalb\'an \& Schatzman \cite{ms00}).
Following the two hypothesis {\bf a)} and {\bf b)} discussed in the
previous section, we show
in panel a)  of Fig.~6 model predictions with an initial Be
abundance equal to the meteoritic value, as assumed in the original works
presenting these models.
In panel b) the curves are instead normalized
to an initial abundance \nbe~$=1.11$ as derived by us
for the young, Li-undepleted
star in our sample; under the assumption that this star
has not depleted any Be, its abundance
is indicative of the initial Be abundance in our abundance scale.

Models including rotational mixing predict
that some Be depletion has
occurred already at 1.7 Gyr and that additional depletion should
take place between 1.7 Gyr and the solar age. 
In addition, at a given age, 
they predict a difference in both \nli~and \nbe~
for stars with the same temperature, but
different initial rotational velocities. The difference
in Li abundance is much larger than the difference in Be abundance. 
In other words, according to these models, some
dispersion in Be should be observed if cluster stars had different
initial rotation rates. A larger dispersion in Li is  predicted.

The absolute Be abundances of M~67 and IC~4651 members shown in panel a)
are, within the errors, in
fairly good agreement with the predictions by rotational mixing.
However, as evidenced in Sect.~4.1, the young cluster member exhibits an
abundance which is a factor $\sim 2$ below meteoritic. The Be content of
this star indeed appears in agreement with the predicted Be abundance 
at an age between 1.7 and 4.5~Gyr, which is clearly inconsistent
with the much younger age of the star.
Panel b) has been made under the assumption that the Be abundance
of the young
star would represent the initial Be abundance; therefore, and by definition, 
this panel does not evidence any inconsistency
for the young cluster member. On the contrary, the observed
Be abundances of both M~67 and IC~4651 members are not fitted by the
models including rotational mixing that predict larger depletion
than observed. Only the warmest star in our sample, S988 in M~67,
is marginally in agreement with model predictions.
Independently from the
initial absolute Be abundance, rotational mixing is not able
to reproduce the {\it relative} Li and Be abundances of our sample stars.
A difference in Be abundance somewhat
larger than our errors is predicted between 1.7 and 4.5~Gyr:
we do not find evidences for such a difference. Furthermore, between
6000 and 5900~K these models predict that a dispersion in 
\nli~of $\sim 0.6$~dex
should correspond to a dispersion in \nbe~of $\sim 0.2$~dex, 
which is
almost a factor of two larger than our errors. Stars S1252 and S1256
in M~67 show 0.75~dex difference in Li, but have virtually the {\it same}
Be abundance, in contrast with the predictions. Even without considering
the derived abundances,
this result is evidenced by the comparison of the spectra of the two
stars in the Be and Li regions shown in Fig.~1.

Gravity waves induced mixing predicts no significant Be depletion at
1.7~Gyr and only a small amount of depletion ($\sim 0.15$~dex) 
for the coolest stars
at the solar age. The predictions of gravity
waves would be in contradiction with our results if hypothesis a) above is true
and the sample stars have undergone some Be depletion; the curves showing
the predictions of these models in panel a) indeed
do not well fit the observed distribution. On the contrary, in panel b)
models including waves provide the best
fit to the observed Be~vs. \teff~diagram of our sample stars.
Mixing driven by waves also seems to reproduce fairly well
the Li vs. temperature distribution of IC~4651 stars, but
is not able to reproduce the Li vs. temperature morphology of
the older M~67 shown in Fig.~7. In particular,
available
models cannot explain the dispersion in Li observed in M~67.
Rotating models including gravity waves are being developed
(Montalb\'an \& Schatzman \cite{ms00}); 
whereas, different rotation rates could provide an explanation for
the dispersion in Li, preliminary qualitative analysis indicates that
high rotation would result in lower predicted Li depletion thus not
reproducing the lower envelope of M~67.

We finally mention that we 
have not taken into account in this discussion Li depletion due to
mass loss. Models including mass loss (Swenson \& Faulkner
\cite{faulk}) predict that Be depletion
starts only when Li has been completely depleted and thus they would be
qualitatively in agreement with our results. Mass loss, however, has
been convincingly shown not to work when applied to Li data alone 
(Swenson \& Faulkner \cite{faulk}).

\section{Conclusions}
We have obtained the first Be measurements of late F/early--G dwarfs in
old open clusters. The Be abundances of old cluster stars are compared with
the abundance of a young Li-undepleted star and with the solar
abundance: we find that all the sample stars share, within the errors
the same Be abundance, irrespective of their age and Li abundance. This
in turn suggests that these stars, including
the Sun, do not undergo any Be depletion during the main sequence,
unless the very unlikely hypothesis is accepted that they deplete Be,
but do not deplete Li in the PMS phases. Our data are thus indicative
of a shallow mixing process that extends deep
enough in the stellar interior for Li burning to occur, but not deep enough to 
destroy Be. At variance with the results for field stars,
we do not find any evidence of correlated Li and Be depletion, which
would support shallow mixing driven by rotation. More in general,
our observations and the comparison with various models
show that the understanding of
the mechanism regulating
the boundary layers between the convective zone and the radiative core
is still not satisfactory and that
the problem of depletion of light elements during the MS
phases of late F/early G--type stars, including our own
Sun, needs further investigation on theoretical grounds. 
None of the models that we have discussed is able to simultaneously fit
the observed Be vs. \teff~and Li vs. \teff~distributions of our sample stars
and in particular their relative amounts of Li and Be depletion.
Our sample stars cover a relatively small interval of effective temperatures;
therefore we cannot extend our results/conclusions to warmer/cooler stars.
Indeed, different mixing mechanisms may be at work in stars much
warmer/cooler than the Sun, with much thinner/thicker convective zones.
In particular, we note that the possibility that the mixing processes
at work in F and G--type stars are different is not unrealistic since
the transition between G- and F-type stars corresponds to the region
in the HR diagram where several changes occur; namely, the surface
convective zones gets thinner in F-type stars, possibly magnetic braking
is not (or less) effective, and stars can retain high rotational
velocities during their permanence on the main sequence.
Most obviously, additional observations should carried out covering
a larger temperature range.

\begin{acknowledgements}
We are grateful to Josefina Montalb\'an for sending us the prediction
of models including gravity waves.
This research was partially supported by MURST grants to SR and RP.
\end{acknowledgements}

{}

\end{document}